# Neutrinos From Particle Decay in the Sun and Earth

Gerard Jungman[a†] and Marc Kamionkowski[b‡]

[a]*Department of Physics, Syracuse University, Syracuse, NY 13244*
[b]*School of Natural Sciences, Institute for Advanced Study, Princeton, NJ 08540*

## ABSTRACT

Weakly interacting massive particles (WIMPs) may be indirectly detected by observation of upward muons induced by energetic neutrinos from annihilation of WIMPs that have accumulated in the Sun and/or Earth. Energetic muon neutrinos come from the decays of $\tau$ leptons, $c$, $b$, and $t$ quarks, gauge bosons, and Higgs bosons produced by WIMP annihilation. We provide analytic expressions, suitable for computing the flux of upward muons, for the neutrino energy spectra from decays of all these particles in the center of the Sun and Earth. These analytic expressions should obviate the need for Monte Carlo calculations of the upward-muon flux. We investigate the effects of polarization of the gauge bosons on the neutrino spectra, and find that they are small. We also present simple expressions for the second moments of the neutrino distributions which can be used to estimate the rates for observation of neutrino-induced muons from WIMP annihilation.

# 1. Introduction

Weakly-interacting massive particles (WIMPs) with masses in the GeV-TeV range are among the leading candidates for the dark matter in our Galactic halo [1]. One of the most promising methods for discovery of WIMPs in the halo is via observation of energetic neutrinos from annihilation of WIMPs in the Sun and/or Earth [2][3][4][5]. WIMPs in the halo can be captured in the Sun and the Earth by elastic scattering from nuclei therein. WIMPs that have accumulated in the Sun or Earth will annihilate into ordinary particles, such as quarks, leptons, and if heavy enough, gauge bosons, Higgs bosons, and top quarks. Many of the leading WIMP candidates—such as the neutralino in supersymmetric theories—are Majorana particles, so they do not annihilate directly into neutrinos [6]. Neutrinos with energies in the multi-GeV range are then produced by decay of the annihilation products, and such neutrinos are potentially observable in many astrophysical neutrino detectors.

The differential energy flux of high-energy neutrinos of type $i$ (e.g., $i = \nu_\mu, \bar\nu_\mu$, etc.) from WIMP annihilation in the Sun (Earth) is

$$\left(\frac{d\phi}{dE}\right)_i = \frac{\Gamma_A}{4\pi R^2} \sum_F B_F \left(\frac{dN}{dE}\right)_{F,i}, \qquad (1.1)$$

where $\Gamma_A$ is the rate of WIMP annihilation in the Sun (Earth), and $R$ is the distance of the Earth from the Sun (or the radius of the Earth). The sum is over all annihilation channels $F$ (e.g., pairs of gauge or Higgs bosons or fermion-antifermion pairs), $B_F$ is the annihilation branch for channel $F$, and $(dN/dE)_{F,i}$ is the differential energy spectrum of neutrino type $i$ at the *surface* of the Sun expected from injection of the particles in channel $F$ in the *core* of the Sun. The total number of neutrinos of type $i$ from final state $F$ is $N_{F,i} = \int (dN/dE)_{F,i} dE$. The spectrum $(dN/dE)_{F,i}$ is a function of the energy of the neutrino and of the energy of the injected particles. As we point out in this paper, the $(dN/dE)_{F,i}$ may also depend on the polarization of the annihilation products.

This paper continues the work of Gaisser, Steigman, and Tilav [4] and Ritz and Seckel [3]. The $dN/dE$ from the $\tau$-lepton and $b$- and $c$-quark final states were first computed analytically using the spectator-quark approximation in Ref. [4], but stopping of heavy hadrons and stopping and absorption of neutrinos in the Sun were neglected. A Monte Carlo calculation which included interactions with the solar medium was performed by Ritz and Seckel [3] (hereafter RS). They also considered decays of the top quark, but those results, obtained assuming a top-quark mass of 60 GeV, are no longer valid. Some of the recent experimental collaborations have followed the example of RS and used Monte Carlo simulations to obtain the neutrino energy spectra [7][8]. However, Monte Carlo calculations can be cumbersome, especially when numerous WIMP candidates are to be explored; analytic expressions are much more convenient. It is also easier to isolate the various interaction effects and theoretical assumptions that enter the calculation with analytic expressions. In this paper, we derive analytic expressions for the neutrino energy distributions from $\tau$-lepton, $c$-, $b$-, and $t$-quark, and gauge- and Higgs-boson decay in the Sun and Earth that can be used in place of Monte Carlo results.

Our results for $b$ and $c$ quarks and $\tau$ leptons are derived from the results of [3] and [4]. Our results for $dN/dE$ from gauge-boson and Higgs-boson final states follow the work of Refs. [9] and [5], but our discussion of the top-quark final state is entirely new. If the top quark is heavier than the $W^\pm$ boson, then it decays almost exclusively into $W$ bosons and $b$ quarks; therefore, the calculation of the neutrino spectrum must be redone. Also note that in many cases where the neutralino is heavier than the top quark, neutralinos annihilate predominantly to the $t\bar t$ final state [10][11][12]. Therefore, evaluation of the neutrino spectrum for the $t\bar t$ final state is crucial to calculation of energetic-neutrino event rates.

Of all our results, the neutrino spectra from injected $b$ and $c$ quarks are subject to the largest theoretical uncertainty. In particular, heavy hadrons will be slowed before they decay as they pass through the Sun [3], and little is known about interactions of heavy hadrons with a surrounding medium. However, this uncertainty enters the Monte Carlo calculations as well, so little accuracy is lost by using analytic results instead. In addition, the effect of heavy-quark stopping becomes small at low injection energies, so the resulting uncertainty is small at low energies. Once the neutralino mass has become large enough that the top-quark annihilation channel opens up, the branching ratio for neutralino annihilation into $b$ quarks becomes negligible. So although the $b$-quark neutrino spectrum may be quite uncertain at high injection energies, the resulting uncertainty in the *total* neutrino event-rate remains small. Note that in all cases, the errors in the upward-muon flux from the neutrino spectra are dwarfed by underlying uncertainty in the local halo density.

When the WIMP is heavy enough that it annihilates primarily to vector bosons and/or top quarks, then the theoretical calculation simplifies greatly. These particles (as well as $\tau$ leptons) do not hadronize and are not slowed before they decay, so relatively simple analytic expressions are obtained for the neutrino spectra from the Earth. The only point at which the analysis becomes somewhat complicated is when considering the stopping of the $b$ quark from top quark decay in the Sun. However, this $b$-quark contribution provides a small fraction of the total neutrino signal from top decay, so the uncertainty introduced from modeling of the hadronic stopping effect is negligible. In the end, we arrive at expressions for $dN/dE$ from the Earth which give results for the upward-muon flux estimated to be accurate to a few percent and results for $dN/dE$ from the Sun which give results accurate to within 5%.

We also calculate the neutrino spectra from polarized gauge bosons. In Ref. [5] it was assumed that the gauge bosons produced by neutralino annihilation were unpolarized. In fact, they are indeed polarized [13]. In this work, we quantify the effect of polarization of gauge bosons from neutralino annihilation on the upward-muon flux from decays of these particles and show that it leads to no more than a 10% correction to the results of Ref. [5].

In most neutrino detectors, the existence of the neutrino is inferred by observation of an upward muon produced by a charged-current interaction in the rock below the detector. The cross section for production of a muon is proportional to the neutrino energy, and the range of the muon in rock is roughly proportional to the muon energy. Therefore, the rate for observation of energetic neutrinos is essentially proportional to the second moment of the neutrino energy spectrum. Inserting the numerical values for the charged-current cross section and the effective range of the muon, and ignoring detector thresholds, the rate per unit detector area for neutrino-induced throughgoing-muon events may be written [3][5]

$$\Gamma_{\text{detect}} = (1.27 \times 10^{-29} \text{ m}^{-2} \text{ yr}^{-1}) \frac{C}{1 \text{ sec}^{-1}} \left(\frac{m_{\tilde{\chi}}}{1 \text{ GeV}}\right)^2 \sum_i a_i b_i \sum_F B_F \langle Nz^2 \rangle_{F,i}, \quad (1.2)$$

for neutrinos from the Sun; the same expression multiplied by $5.6 \times 10^8$ (the square of the ratio of the Earth-Sun distance to the Earth's radius) gives the rate for neutrino events from the Earth. Here, $C$ is the WIMP capture rate in the Sun or Earth, $B_F$ is the branching fraction for WIMP annihilation to the final state $F$, the sum on $i$ is over muon neutrinos and muon anti-neutrinos, the $a_i$ are neutrino scattering coefficients, $a_\nu = 6.8$ and $a_{\bar{\nu}} = 3.1$, and the $b_i$ are muon range coefficients, $b_\nu = 0.51$ and $b_{\bar{\nu}} = 0.67$. The quantity

$$\langle Nz^2 \rangle_{F,i}(E_{in}) \equiv \frac{1}{E_{in}^2} \int \left(\frac{dN}{dE}\right)_{F,i}(E_\nu, E_{in}) E_\nu^2 \, dE_\nu, \quad (1.3)$$

is the second moment of the spectrum of neutrino type $i$ from final state $F$ scaled by the square of the injection energy $E_i$ of the annihilation products. The quantity $z = E_\nu/E_{in}$ is the neutrino energy scaled by the injection energy.

Strictly speaking, neutrino telescopes observe neutrinos only with energies above a given threshold. Therefore, the event rate is proportional to the contribution to the second moment from neutrinos with energies above threshold—that is, there is a lower bound to the integral in Eq. (1.3). To obtain the most accurate experimental information, a detailed calculation of the neutrino spectra should be folded in with the detector response. For example, an energetic-neutrino signal from the Sun must be distinguishable from the atmospheric-neutrino background. The direction of the neutrino-induced muon is correlated with the parent-neutrino direction only within an angular window of roughly $\sqrt{\text{GeV}/E_\nu}$ radians. Therefore, a proper determination of the signal-to-noise ratio requires knowledge of the neutrino energy spectrum.

On the other hand, in cases where the WIMP is quite massive, most of the neutrinos have energies large enough to produce a muon with an energy much higher than the detector threshold. For example, if a 100-GeV WIMP annihilates to gauge bosons, the typical neutrino energy is roughly half that, and the typical muon energy typically half that, about 25 GeV. Typical thresholds for current and next-generation detectors are no more than 10 GeV, so the vast majority of the neutrinos in this example are above threshold. In addition, even if a non-negligible fraction of the neutrinos have energies below the detector threshold, the total second moment of the energy distribution is still primarily determined by the higher-energy neutrinos. The contribution of neutrinos with energies below threshold to the upward-muon flux will be $\mathcal{O}(E_{\text{thresh}}^3/m_\chi^3)$ where $E_{\text{thresh}}$ is the threshold energy. Therefore, in cases where the WIMP mass is significantly larger than the detector threshold, the expression for the rate for neutrino-induced upward muons, Eq. (1.2), together with the results for $\langle Nz^2 \rangle$ presented here will provide a good theoretical estimate to compare with experimental determinations of the flux of upward muons from the Sun and/or Earth.

The most promising method of detection of the energetic neutrinos, especially for higher-mass WIMPs, is the upward-muon signal. Therefore, we calculate the energy spectrum of muon neutrinos and antineutrinos only.

In the next Section, we discuss some general results on neutrino spectra. We relate the neutrino spectrum from decay of a moving particle to the neutrino spectrum from decay of the particle at rest, and we show how the neutrino spectra from particle decays in the Sun are obtained from those from decay in the Earth. We discuss how hadronization and stopping of $b$ and $c$ quarks affect the neutrino spectra. We also discuss how neutrino spectra from Higgs bosons are obtained from the neutrino spectra from the Higgs-boson decay products. Much of this is review, but useful for the discussion in the Sections that follow. In Section 3, we calculate the neutrino spectra from decays of $b$ and $c$ quarks and $\tau$ leptons. In Section 4, we discuss the neutrino spectra from decays of gauge bosons and quantify the effect of gauge-boson polarization on the neutrino spectra. In Section 5, we calculate the neutrino spectra from decays of top quarks, and we summarize our results in Section 6.

## 2. Preliminaries

For all of the particles considered, calculation of the neutrino distribution in the rest frame of the decaying particle, $(dN/dE)^{\text{rest}}(E_\nu)$, as a function of the neutrino energy $E_\nu$, is a standard exercise. Once the rest-frame distribution is known, the energy distribution of a particle moving with an energy $E_d$, velocity $\beta$, and $\gamma = (1-\beta^2)^{-1/2} = E_d/m_d$ (where $m_d$ is the decaying-particle mass) is related to the rest-frame distribution by

$$\left(\frac{dN}{dE}\right)^\oplus (E_d, E_\nu) = \frac{1}{2} \int_{E_-(E)}^{E_+(E)} \frac{d\epsilon}{\epsilon} \frac{1}{\gamma\beta} \left(\frac{dN}{dE}\right)^{\text{rest}} (\epsilon), \qquad (2.1)$$

where

$$E_\pm(E) = \frac{E_\nu}{\gamma(1 \mp \beta)}. \qquad (2.2)$$

This assumes that there is no prefered direction in the rest frame of the parent particle. In particular, it applies to decay of scalar or unpolarized particles. If the decaying particle is polarized and the neutrino direction is correlated with the spin of the decaying particle, Eq. (2.1) does not apply. In almost all cases we consider, the decaying particles are unpolarized, and we show that in the few cases where they are polarized, the effect of polarization is small or can be accounted for relatively easily.

The Earth is thin enough that stopping of heavy hadrons and stopping and absorption of neutrinos as they pass through the Earth can be neglected. Therefore, if the rest-frame distribution from a particular decaying particle is known, Eq. (2.1) gives the correct $dN/dE$ for particle decay in the Earth; hence the superscript $\oplus$ in Eq. (2.1). This also implies that the neutrino and antineutrino spectra from injection of a given particle-antiparticle pair in the center of the Sun are the same. Also, if the scaled second moment in the rest frame of the decaying particle, $\langle Nz^2 \rangle^{\text{rest}}$, is known, then the scaled second moment for a particle that decays with a velocity $\beta$ in the Earth is simply

$$\langle Nz^2 \rangle(\beta) = \langle Nz^2 \rangle^{\text{rest}} (1 + \frac{\beta^2}{3}). \qquad (2.3)$$

On the other hand, energetic neutrinos will lose energy via neutral-current interactions with the solar medium and become absorbed via charged-current interactions as they pass through the Sun. To account for these effects, we use the calculation of RS that a neutrino injected with an energy $E$ leaves the Sun with energy

$$E_f = \frac{E}{1 + E\tau_i}, \qquad (2.4)$$

where $\tau_\nu = 1.01 \times 10^{-3}$ GeV$^{-1}$ and $\tau_{\bar{\nu}} = 3.8 \times 10^{-4}$ GeV$^{-1}$, and with probability

$$P_f = \left(\frac{1}{1 + E\tau_i}\right)^{\alpha_i}, \qquad (2.5)$$

where $\alpha_\nu = 5.1$ and $\alpha_{\bar{\nu}} = 9.0$. As a result, the neutrino spectrum for a particle decaying with energy $E_d$ in the Sun, $(dN/dE)^\odot$, is related to the neutrino spectrum for a particle decaying with energy $E_d$ in the Earth, $(dN/dE)^\oplus$, by

$$\left(\frac{dN}{dE}\right)^\odot_i (E_d, E_\nu) = (1 - E_\nu\tau_i)^{\alpha_i - 2} \left(\frac{dN}{dE}\right)^\oplus (E_d, E_m), \qquad (2.6)$$

where $E_m = E_\nu/(1 - E_\nu\tau_i)$ is the energy a neutrino had at the core of the Sun if it exits with energy $E_\nu$. Note that stopping and absorption are different for neutrinos and antineutrinos, so the spectrum of neutrinos from WIMP annihilation in the Sun is different than the spectrum of antineutrinos. The spectrum of neutrinos is the same as that for antineutrinos for decays in the Earth, so we neglect the subscript $i$ on neutrino spectra from the Earth.

If a $b$ or $c$ quark is injected into the center of the Earth, it will lose energy during hadronization. As a result, the energy at which the hadron decays in the Earth, $E_d$, is related to the energy $E_i$ at which it is injected by $E_d = z_f E_i$. We are interested primarily in the second moment of the neutrino distributions, so we use the square root of the second moments of the fragmentation functions computed by RS—that is, we take $z_f = 0.58$ for $c$ quarks and $z_f = 0.73$ for $b$ quarks. Actually, the distribution of $z_f$ is described by a fragmentation function [14], but this fragmentation function is highly localized around $z_f$. Comparison of higher order moments of $z_f$ suggests that no more than $\mathcal{O}(5\%)$ accuracy is lost by using the central value of the fragmentation function.

In addition, the core of the Sun (but *not* the Earth) is dense enough that $b$- and $c$-quark hadrons will interact with the solar medium and be slowed appreciably before they decay [3]. If a hadron initially has an energy $E_0 = z_f E_i$ (after hadronization of a quark injected with energy $E_i$), then it will decay with an energy $E_d$ picked from a decay distribution [3],

$$\left(\frac{1}{N}\frac{dN}{dE_d}\right)^{\text{hadron}} (E_0, E_d) = \frac{E_c}{E_d^2} \exp\left[E_c\left(\frac{1}{E_c} - \frac{1}{E_d}\right)\right], \qquad (2.7)$$

where $E_c = 250$ GeV for $c$ quarks, $E_c = 470$ GeV for $b$ quarks.

The $n$th moment of this distribution, for a given initial energy $E_0$, is

$$\langle E_d^n \rangle (E_0) = \int_0^{E_0} \left(\frac{1}{N}\frac{dN}{dE_d}\right)^{\text{hadron}} (E_0, E) E^n \, dE, \qquad (2.8)$$

so in particular, the average energy at which a hadron decays is

$$\langle E_d \rangle (E_0) = E_c \exp\left(\frac{E_c}{E_0}\right) E_1\left(\frac{E_c}{E_0}\right), \qquad (2.9)$$

where $E_1(x) = \int_x^\infty (e^{-y}/y) dy$ [15], and the rms value of the decay energy is

$$E_d^{\text{rms}}(E_0) = \sqrt{\langle E_d^2 \rangle} = \sqrt{E_c(E_0 - \langle E_d \rangle)}. \qquad (2.10)$$

These quantities will come in handy in the following Section.

Of all the factors that enter into the calculation of the neutrino spectrum, stopping of heavy hadrons is the most uncertain theoretically; very little is known about the interaction of heavy-quark hadrons with dense matter. The functional form of the decay distribution, Eq. (2.7), is physically well-motivated, but the values of $E_c$ listed here are subject to some (perhaps sizeable) uncertainty. However, this theoretical uncertainty also enters the Monte Carlo calculation; in this regard, Monte Carlo simulations offer no improvement over the analytic result. Also, the effect of hadron stopping is small at lower energies, so the uncertainty introduced into the neutrino spectra is relatively small. Stopping becomes much stronger—and the subsequent theoretical uncertainty much larger—at higher injection energies, but in most cases where the WIMP is massive enough that it can annihilate into top quarks, it annihilates almost exclusively to top quarks, gauge bosons, and/or Higgs bosons [10][11][12]. Therefore, the uncertainty in the *total* neutrino spectrum due to poor understanding of hadron stopping is never very large.

Before continuing, we remind the reader that light ($u$, $d$, and $s$) quarks will produce long-lived hadrons that come to rest in the core of the Sun or Earth before they decay; therefore, these final states produce no energetic neutrinos. The same is true of neutrinos from muon decay. On the other hand, $\tau$ leptons, top quarks, and gauge and Higgs bosons will decay essentially immediately, so there is *no* energy loss due to slowing or hadronization before they decay.

Finally, consider a chain decay process, $H \to l_1 l_2$, where a heavy particle, $H$, decays to two lighter particles, $l_1$ and $l_2$, which then decay to energetic neutrinos. For example, the heavy particle could be one of the Higgs bosons which then decays to two light fermions. In the rest frame of the decaying $H$, the energies of the two light particles are $E_1 = (m_H^2 + m_1^2 - m_2^2)/(2m_H)$ and $E_2 = (m_H^2 + m_2^3 - m_1^2)/(2m_H)$, and their velocities are $\beta_1 = (1 - m_1^2/E_1^2)^{1/2}$ and $\beta_2 = (1 - m_2^2/E_2^2)^{1/2}$, where $m_H$, $m_1$, and $m_2$ are the masses of $H$, $l_1$, and $l_2$, respectively. Suppose the neutrino spectrum from decay of $l_1$ as a function of neutrino energy $E_\nu$ for a given injection energy $E_i$ is $(dN/dE)_1(E_i, E_\nu)$, and similarly for neutrinos from decays of $l_2$. Then the spectrum from the heavy particle $H$, which decays while moving with a velocity $\beta_H$ and energy $E_H = \gamma_H m_H$ is

$$\left(\frac{dN}{dE}\right)_H (E_H, E_\nu) = \sum_{f=1,2} \frac{1}{2\gamma_H E_f \beta_f \beta_H} \int_{\gamma_H E_f (1-\beta_H\beta_f)}^{\gamma_H E_f (1+\beta_H\beta_f)} \left(\frac{dN}{dE}\right)_f (E, E_\nu) \, dE, \quad (2.11)$$

where the sum is over the two decay particles, and $E_f$ and $\beta_f$ are the decay-particle energies and velocities in the rest frame of the decaying $H$. Also, if the scaled second moments of the neutrino distribution, $\langle Nz^2 \rangle_1 (E_i)$ and $\langle Nz^2 \rangle_2 (E_i)$, from decays of $l_1$ and $l_2$, respectively, are known for injection energies $E_i$, then the $\langle Nz^2 \rangle$ from decay of $H \to l_1 l_2$ is [16]

$$\langle Nz^2 \rangle_H (E_H) = \frac{1}{E_H^2} \sum_{f=1,2} \frac{1}{2\gamma_H E_f \beta_f \beta_H} \int_{\gamma_H E_f (1-\beta_H\beta_f)}^{\gamma_H E_f (1+\beta_H\beta_f)} E^2 \langle Nz^2 \rangle_f (E) \, dE. \quad (2.12)$$

These equations can be used, for example, to determine the neutrino spectra from Higgs-boson decay. Higgs bosons do not decay directly to neutrinos, so all the energetic-neutrino signal comes from the decay of the Higgs decay products. Given the neutrino spectra from the Higgs decay channels (e.g., quarks, leptons, and if heavy enough, gauge bosons and/or top quarks) presented below, the neutrino spectrum and/or $\langle Nz^2 \rangle$ from Higgs decay are obtained by summing Eqs. (2.11) and/or (2.12) over all decay channels of the Higgs with the appropriate branching ratios. We will use these results below to obtain the $\langle Nz^2 \rangle$ from top-quark decay. Also, note that these relations, Eqs. (2.11) and (2.12), are valid for neutrino spectra from particle decay in both the Sun and the Earth.

Given the results in this Section, the neutrino spectra from each final state follow from the kinematics of two- and three-body decay, and we will discuss each final state below.

## 3. Neutrinos From $\tau$ Leptons and $b$ and $c$ Quarks

### 3.1. Neutrinos From $\tau$ Leptons

We begin with the neutrino energy spectra from $\tau$ leptons. Calculation of these spectra is easier than that for $b$ and $c$ quarks, since $\tau$ leptons do not hadronize, and they are not slowed before stopping. In addition, we consider decay of unpolarized $\tau$ leptons only. For neutralinos, this is valid because the $\tau$ leptons produced by neutralino annihilation are unpolarized [13]. If for some unforeseen reason, the injected $\tau$ leptons are polarized, the results for the neutrino spectra will be slightly different.

In the rest frame of the $\tau$, the energy distribution of the muon neutrino from the decay $\tau \to \mu \bar{\nu}_\mu \nu_\tau$ is [17]

$$\left(\frac{dN}{dE_\nu}\right)^{\text{rest}}_{\bar{\tau}\tau} = \frac{96 \Gamma_{\tau \to \mu\nu\nu}}{m_\tau^4} E_\nu^2 (m_\tau - 2E_\nu), \qquad 0 \le E_\nu \le \frac{1}{2} m_\tau, \quad (3.1)$$

where $\Gamma_{\tau \to \mu\nu\nu} \simeq 0.18$ is the branching ratio for $\tau$ decay to muons. The spectrum of neutrinos from decays of $\tau$ leptons moving with an energy $E_i$ (the injection energy), velocity $\beta$, and $\gamma = (1 - \beta^2)^{-1/2}$ is

$$\left(\frac{dN}{dE_\nu}\right)^{\oplus}_{\bar\tau\tau}(E_i, E_\nu) = \frac{48\Gamma_{\tau \to \mu\nu\nu}}{\beta\gamma m_\tau^4}\left(\frac{1}{2}m_\tau E_\nu^2 - \frac{2}{3}E_\nu^3\right)^{\min(m_\tau/2, E_+)}_{E_-}, \quad (3.2)$$

where $E_\pm = E_\nu \gamma^{-1}(1 \mp \beta)^{-1}$. In the relativistic limit ($\beta \to 1$), this becomes $(dN/dE) = (2\Gamma/E_i)(1 - 3x^2 + 2x^3)$, for $0 \leq E_\nu \leq E_i$, where $x = E_\nu/E_i$. Given the neutrino distribution from $\tau$ decay in the Earth, the spectrum from $\tau$ decay in the Sun is obtained by using Eq. (2.6) to account for stopping and absorption of neutrinos.

Applying the results of Section 2, the scaled second moment from injection of $\bar\tau\tau$ pairs with velocity $\beta$ in the core of the Earth is

$$\langle Nz^2 \rangle^{\oplus}_{\bar\tau\tau} = \frac{\Gamma_{\tau \to \mu\nu\nu}}{10}(1 + \frac{\beta^2}{3}). \quad (3.3)$$

The integral over the distributions from the Sun can also be performed, and the result is

$$\langle Nz^2 \rangle^{\odot}_{\bar\tau\tau,i}(E_i) = \Gamma_{\tau \to \mu\nu\nu}\, h_{\tau,i}(E_i\tau_i), \quad (3.4)$$

where the $\tau_i$ are the neutrino stopping coefficients given in Section 2. For muon neutrinos, the function $h_{\tau,i}(y)$ is given by

$$h_{\tau,\nu_\mu}(y) = \frac{1}{30}\frac{4+y}{(1+y)^4}, \quad (3.5)$$

and for antineutrinos, the appropriate function is

$$h_{\tau,\bar\nu_\mu}(y) = \frac{1}{1260}\frac{168 + 354y + 348y^2 + 190y^3 + 56y^4 + 7y^5}{(1+y)^8}. \quad (3.6)$$

In deriving Eq. (3.5), we took $\alpha_{\nu_\mu} = 5.0$, which is an excellent approximation to the RS value of $\alpha_{\nu_\mu} = 5.1$. In addition, the $h$ functions were obtained using the $\beta \to 1$ limit of $(dN/dE)^{\oplus}_{\bar\tau\tau}$. If $\beta$ is not near unity, then $E_i$ must be close to $m_\tau$, in which case neutrino absorption and stopping will be negligible and the $dN/dE$ from the Earth can be used. Also, in the low-energy limit ($E_i \ll \tau_i^{-1} \simeq$ TeV), simple but accurate expressions can be obtained from Taylor expansions of the $h$ functions.

In addition to the energetic muons produced by direct decays of the $\tau$, there will be additional indirect energetic muon neutrinos from the $\tau$ neutrinos. The $\tau$ neutrino from the $\tau$ decay may produce an additional $\tau$ lepton via a charged-current interaction with the solar medium as it passes through the Sun, and this $\tau$ lepton will then decay and produce an additional muon neutrino [3]. The energy of this muon neutrino is much smaller than that from direct decay of the original $\tau$ lepton. Although these neutrinos may contribute non-negligibly to $dN/dE$ at low neutrino energies, this contribution to the upward-muon signal is tiny, so it is justifiably ignored when considering upward muons. Therefore, Eqs. (3.3), (3.4), (3.5), and (3.6) account for the overwhelming majority of the energetic neutrinos that give rise to upward muons. In addition, they correctly take into account all the important physical processes: the $\tau$-decay kinematics, and stopping and absorption of neutrinos as they pass through the Sun. Therefore, these expressions are subject to little theoretical uncertainty and can be used in place of the results of Monte Carlo calculations.

### 3.2. Neutrinos from b and c Quarks

The three-body decay of $b$- and $c$-quark hadrons is similar to that for $\tau$ decay, and the treatment of neutrino stopping and absorption in the Sun is similar; however, calculation of the neutrino spectra is slightly more complicated due to hadronization of the quarks and stopping of the heavy hadrons in the Sun. Still, we can treat these effects analytically without resorting to Monte Carlos. Hadronization is simple: the energy of a quark is degraded from its injection energy $E_i$ to the hadron energy $E_0 = z_f E_i$. Stopping is straightforward as well: we must integrate the neutrino energy spectrum from a hadron which decays with energy $E_d$ over the distribution of decay energies, Eq. (2.7).

In the rest frame of the $b$-quark hadron, the energy distribution of the neutrino from the decay $b \to c\mu\nu$ is (neglecting the small correction due to finite $c$-quark mass) similar to that from $\tau$-lepton decay [17]:

$$\left(\frac{dN}{dE_\nu}\right)^{\text{rest}}_{b\bar b} = \frac{96\Gamma_{b \to \mu\nu X}}{m_b^4}E_\nu^2(m_b - 2E_\nu), \qquad 0 \leq E_\nu \leq \frac{1}{2}m_b, \quad (3.7)$$

where $\Gamma_{b\to\mu\nu X} \simeq 0.103$ is the branching ratio for inclusive semileptonic decay of the $b$ quark into muons [18]. A $b$ quark injected with energy $E_i$ decays in the Earth with an energy $E_d = z_f E_i$, velocity $\beta$, and $\gamma = (1-\beta^2)^{-1/2} = E_d/m_b$ and the resulting neutrino spectrum is

$$\left(\frac{dN}{dE_\nu}\right)^{\oplus}_{b\bar{b}} = \frac{48\Gamma_{b\to\mu\nu X}}{\beta\gamma m_b^4}\left(\frac{1}{2}m_b E_\nu^2 - \frac{2}{3}E_\nu^3\right)^{\min(m_b/2, E_+)}_{E_-}, \qquad (3.8)$$

where $E_\pm = E_\nu \gamma^{-1}(1\mp\beta)^{-1}$. In the relativistic limit ($\beta\to 1$), this becomes $(dN/dE) = (2\Gamma/E_d)(1 - 3x^2 + 2x^3)$, for $0\le E_\nu \le E_d$, where $x = E_\nu/E_d$.

The neutrino spectrum from $b$ decay in the Sun is obtained by applying Eq. (2.6) to Eq. (3.8), and integrating over the distribution of hadron decay energies, Eq. (2.7):

$$\left(\frac{dN}{dE}\right)^{\odot}_{b\bar{b},i}(E_i, E_\nu) = \int_0^{z_f E_i}\left(\frac{1}{N}\frac{dN}{dE_d}\right)^{\text{hadron}}(E_d)$$
$$\times (1 - E_\nu\tau_i)^{\alpha_i - 2}\left(\frac{dN}{dE}\right)^{\oplus}_{b\bar{b}}(E_d, E_m)\, dE_d$$
$$= \frac{2\Gamma_{b\to\mu\nu X}}{E_c}(1 - E_\nu\tau_i)^{\alpha_i - 2}\exp\left(\frac{E_c}{E_0} - x\right)$$
$$\times (1 - 18y^2 + 24x^2 y^3 + x + 48y^3 + 8y^3 x^3 - 9y^2 x^2$$
$$+ 48y^3 x + 2y^3 x^4 - 18y^2 x - 3y^2 x^3), \qquad (3.9)$$

where $E_m = E_\nu/(1 - E_\nu\tau_i)$, $x = \max(E_c/E_0, E_c/E_\nu)$, $E_0 = z_f E_i$, and here, $y = E_\nu/E_c$. The value of $E_c$ to be used here is 470 GeV. Eq. (3.9) was obtained using the $\beta\to 1$ limit of $(dN/dE)^{\oplus}_{b\bar{b}}$, so it is not strictly valid for $E_i$ near $m_b$. On the other hand, if the injection energy is so low that $\beta$ is not near unity, interactions with the solar medium are small and the spectrum from decay in the Earth can be used.

In addition to the prompt neutrinos from $b$-quark decay, there will be additional energetic neutrinos from subsequent decays of the $c$ quark. However, the typical energy of these neutrinos is roughly $1/3$ the energy of the neutrinos from prompt decay, so they contribute only about 10% to the total upward-muon signal. Since neutralino annihilation into $b$ quarks is always accompanied by annihilation into $\tau$ leptons, $c$ quarks, and possibly top quarks and gauge/Higgs bosons, the error introduced into the upward-muon rate by including only prompt neutrinos from $b$ decay will never be greater than 10%. If greater accuracy is desired, the contribution from $c$ quarks can be included using the $c$-quark neutrino spectra below and the chain-decay equations of Section 2.

The neutrino spectra from charmed hadrons are similar, but the three-body decay kinematics is slightly different. In the rest frame of the charmed hadron, the energy distribution of the neutrino from the decay $c\to s\mu\nu$ is (neglecting the small correction due to finite strange-quark mass) [17]

$$\left(\frac{dN}{dE_\nu}\right)^{\text{rest}}_{c\bar{c}} = \frac{16\Gamma_{c\to\mu\nu X}}{m_c^4}E_\nu^2(3m_c - 4E_\nu), \qquad 0 \le E_\nu \le \frac{1}{2}m_c, \qquad (3.10)$$

where $\Gamma_{c\to\mu\nu X} \simeq 0.13$ is the branching ratio for inclusive semileptonic decay of the $c$ quark into muons. (We obtain this value for $\Gamma_{c\to\mu\nu X}$ by assuming charged and neutral charmed mesons are produced with equal probability and averaging the branching ratios for inclusive semileptonic decays to muons [18]. Our value for this branching ratio appears to be almost twice as large as that used by RS, so our neutrino spectra from $c$-quark decay will be significantly larger than those obtained by RS.) A $c$ quark injected with energy $E_i$ decays in the Earth with an energy $E_d = z_f E_i$, velocity $\beta$, and $\gamma = (1-\beta^2)^{-1/2} = E_d/m_c$ and the resulting neutrino spectrum is

$$\left(\frac{dN}{dE_\nu}\right)^{\oplus}_{c\bar{c}} = \frac{8\Gamma_{c\to\mu\nu X}}{\beta\gamma m_c^4}\left(\frac{3}{2}m_c E_\nu^2 - \frac{4}{3}E_\nu^3\right)^{\min(m_c/2, E_+)}_{E_-}, \qquad (3.11)$$

where $E_\pm = E_\nu \gamma^{-1}(1\mp\beta)^{-1}$. In the relativistic limit ($\beta\to 1$), this becomes $(dN/dE) = (\Gamma/E_d)[(5/3) - 3x^2 + (4/3)x^3]$, for $0\le E_\nu \le E_d$, where $x = E_\nu/E_d$.

The derivation of the neutrino spectrum from $c$ decay in the Sun is similar to that for $b$ quarks above. The result is

$$\left(\frac{dN}{dE}\right)^{\odot}_{c\bar{c},i}(E_i, E_\nu) = \frac{\Gamma_{c\to\mu\nu X}}{3E_c}(1 - E_\nu\tau_i)^{\alpha_i - 2}\exp\left(\frac{E_c}{E_0} - x\right)$$
$$\times (5 - 54y^2 + 48y^3 x^2 + 5x + 96y^3 + 16y^3 x^3 - 27x^2 y^2$$
$$+ 96y^3 x + 4y^3 x^4 - 54y^2 x - 96y^2 x^3), \qquad (3.12)$$

where as before, $E_m = E_\nu/(1 - E_\nu \tau_i)$, $x = \max(E_c/E_0, E_c/E_\nu)$, $E_0 = z_f E_i$, and here, $y = E_\nu/E_c$. The value of $E_c$ to be used here is 250 GeV.

Now we list expressions for the $\langle Nz^2 \rangle$. The expression for decay of a $b$ quark moving with velocity $\beta$ in the Earth is

$$\langle Nz^2 \rangle^\oplus_{b\bar{b}} = \frac{z_f^2 \Gamma_{b \to \mu\nu X}}{10}(1 + \frac{\beta^2}{3}), \quad (3.13)$$

and that from $c$-quark decay is

$$\langle Nz^2 \rangle^\oplus_{c\bar{c}} = \frac{2z_f^2 \Gamma_{c \to \mu\nu X}}{15}(1 + \frac{\beta^2}{3}). \quad (3.14)$$

The second moments from $b$- and $c$-quark decay in the Sun can be obtained by integrating the correct expressions for $dN/dE$. However, it turns out to be easier to switch the order of integration—that is, first compute $\langle Nz^2 \rangle$ for a given decay energy, $E_d$, and then integrate over the hadronic decay-energy distribution. The result is

$$\langle Nz^2 \rangle_{f,i}(E_i) = \frac{\Gamma_{f \to \mu\nu X}}{E_i^2} E_c^2 e^{E_c/E_0} \int_{E_c/E_0}^\infty \frac{dx}{x^2} e^{-x} h_{f,i}(E_c \tau_i/x)$$
$$\simeq \frac{\langle E_d \rangle^2}{E_i^2} h_{f,i}\left(\sqrt{\langle E_d^2 \rangle} \tau_i\right), \quad (3.15)$$

where the subscript $f$ denotes $b$ or $c$ quarks, and the subscript $i$ refers to neutrino type ($\nu_\mu$ or $\bar{\nu}_\mu$), and the moments $\langle E_d^n \rangle$ are those that were given in Section 2. The integral in the first line cannot be performed analytically, but we have found that the approximation in the second line is accurate to a few percent for injection energies less than a few TeV. For $b$ quarks, the functions $h_{b,i}$ are the same as those for $\tau$ leptons given in the previous subsection: $h_{b,i}(y) = h_{\tau,i}(y)$ [c.f., Eqs. (3.5) and (3.6)]. For neutrinos, the $c$-quark functions are,

$$h_{c,\nu_\mu}(y) = \frac{1}{180} \frac{32 + 25y + 5y^2}{(1+y)^5}, \quad (3.16)$$

and for antineutrinos,

$$h_{c,\bar{\nu}_\mu}(y) = \frac{1}{7560}(1344 + 3186y + 3834y^2 \\ + 2786y^3 + 1242y^4 + 315y^5 + 35y^6)/(1+y)^9. \quad (3.17)$$

Eqs. (3.13), (3.14), and (3.15), with the expressions for the appropriate $h(y)$ functions, correctly describe the second moments of the neutrino distributions from $b$- and $c$-quark decay in the Sun and Earth and include the effects of hadronization and stopping of heavy quarks and stopping and absorption of neutrinos in the Sun. They therefore can be used in place of Monte Carlo calculations.

## 4. Neutrinos From Gauge Bosons

A $W$ boson decays directly to a muon and a muon neutrino 10.5% of the time, and a $Z$ decays to a muon neutrino-antineutrino pair 6.7% of the time. Additional muon neutrinos are indirectly produced by $W$ decay to $b$ and $c$ quarks and $\tau$ leptons. The energies of these neutrinos are generally smaller, so the contribution provided by these indirect neutrinos to $\langle Nz^2 \rangle$ is a small fraction of that from the direct neutrinos. Recall that light-fermion final states produce only very low energy neutrinos since they become thermalized before decaying [3], and their contributions to the neutrino energy moments are completely negligible.

We will first calculate the $\langle Nz^2 \rangle$ from decay of polarized and unpolarized gauge bosons in the Earth, and show that polarization is never more than a 10% effect. Given these results, we will present expressions for $dN/dE$ from decay of unpolarized gauge bosons only in the Sun and in the Earth, and then we will provide an analytic result for $\langle Nz^2 \rangle$ for gauge-boson decay in the Sun.

In the rest frame of the vector boson, the neutrino is emitted with an energy equal to half the vector-boson mass; in the laboratory frame, the energy of the emitted neutrino is $E_\nu = E_i(1 + \beta \cos\theta)/2$ where $\theta$ is the angle in the rest frame between the neutrino direction and the direction of motion of the center of mass, $E_i$ is the energy at which the vector boson is injected, and $\beta$ is its velocity. Therefore, for a vector boson injected into the Earth with energy $E_i$, the second moment of the resulting neutrino spectrum is

$$\langle E_\nu^2 \rangle = \frac{E_i^2}{4} \frac{\int P(\cos\theta)(1 + \beta \cos\theta)^2 \, d(\cos\theta)}{\int P(\cos\theta) \, d(\cos\theta)}, \quad (4.1)$$

where $P(\cos\theta)$ is the decay angular distribution in the rest frame of the gauge boson, and and $\Gamma_\nu$ is the fraction of vector-boson decays which produce a neutrino.

For unpolarized gauge bosons, the decay is isotropic, and [5]

$$\langle E_\nu^2 \rangle = \frac{E_i^2}{4}\left(1 + \frac{\beta^2}{3}\right). \qquad (4.2)$$

If the gauge boson has helicity $h = \pm 1$ (transversely polarized), then $P(\cos\theta) \propto (1 \pm \cos\theta)^2$, and

$$\langle E_\nu^2 \rangle_T = \frac{E_i^2}{4}\left(1 + \frac{2\beta^2}{5} \pm 5\beta\right), \qquad (4.3)$$

and if the gauge boson is longitudinally polarized ($h = 0$), then $P(\cos\theta) \propto \sin^2\theta$ and

$$\langle E_\nu^2 \rangle_L = \frac{E_i^2}{4}\left(1 + \frac{\beta^2}{5}\right). \qquad (4.4)$$

The velocities of WIMPs in the Sun or halo are $v \ll 1$, so annihilation proceeds only through the lowest angular-momentum state, the $s$ wave. Although the $s$-wave cross sections for annihilation of neutralinos into $t\bar{t}$ [19][20][21], gauge-boson pairs [10], and Higgs-boson–gauge-boson pairs [5] were calculated previously, the helicity-amplitude formalism used by Drees and Nojiri [13] is needed to describe the polarization state of the annihilation products. According to the results of [13], the gauge bosons produced in the $s$-wave annihilation processes $\tilde{\chi}\tilde{\chi} \to W^+W^-$ and $\tilde{\chi}\tilde{\chi} \to ZZ$ are produced with a tensor polarization; that is, the gauge bosons are transversely polarized, half with spin aligned ($h = 1$) with the direction of motion and half with spins antialigned ($h = -1$). Therefore, the scaled second moment of the neutrino spectrum from neutralino annihilation into $W^+W^-$ pairs in the Earth is an average of that from the two transverse-polarization states. The second moment for neutrinos from the Earth is then given by

$$\langle Nz^2 \rangle^{\oplus}_{\tilde{\chi}\tilde{\chi}\to WW} = \Gamma_{W\to\mu\nu}\frac{1}{4}(1 + \frac{2}{5}\beta_W^2), \qquad (4.5)$$

where $\Gamma_{W\to\mu\nu} = 0.105$. There will be additional muon neutrinos from $\tau$ leptons from $W$ decay, but their contribution to $\langle Nz^2 \rangle$ is small. Only one in five $\tau$ decays produces a muon neutrino, and the energy of this neutrino is typically only one third the energy of a neutrino from direct decay. Therefore, Eq. (4.5) properly accounts for polarization and should be accurate to better than 3%.

The expression for $ZZ$ pairs is obtained similarly to the above. We include an overall factor of two, which counts the $Z$ bosons, and arrive at the result

$$\langle Nz^2 \rangle^{\oplus}_{\tilde{\chi}\tilde{\chi}\to ZZ} = 2\Gamma_{Z\to\nu_\mu\bar{\nu}_\mu}\frac{1}{4}(1 + \frac{2}{5}\beta_Z^2), \qquad (4.6)$$

where $\Gamma_{Z\to\nu_\mu\bar{\nu}_\mu} = 0.067$. Again, there will be additional neutrinos from $Z$ decay to $b\bar{b}$, and to a lesser extent, $c\bar{c}$ and $\bar{\tau}\tau$, and the magnitude of their contribution to $\langle Nz^2 \rangle$ is easily estimated. Although the branching ratio to $b\bar{b}$ is 2.2 times that into $\nu_\mu\bar{\nu}_\mu$, a muon neutrino is produced in only one in ten $b$-quark decays. In addition, the $b$-quark energy is reduced to 0.7 times its original energy during hadronization, and the resulting muon neutrino carries only a third of that energy. By performing similar estimates for the $\tau$ lepton and $c$ quark, we find that Eq. (4.6) is accurate to better than 4%.

Gauge bosons that are produced in the annihilation processes $\tilde{\chi}\tilde{\chi} \to ZH$, where $H$ is one of the two neutral scalar Higgs bosons, or $\tilde{\chi}\tilde{\chi} \to W^+H^-$ (or its charge conjugate), where $H^\pm$ is the charged Higgs boson, are longitudinally polarized [13], so Eq. (4.4) is the appropriate result to use for these final states. Thus we have

$$\langle Nz^2 \rangle^{\oplus}_{\tilde{\chi}\tilde{\chi}\to ZH} = \Gamma_{Z\to\nu_\mu\bar{\nu}_\mu}\frac{1}{4}(1 + \frac{1}{5}\beta_Z^2) + \text{Higgs decay contribution}. \qquad (4.7)$$

The Higgs-decay contribution can be evaluated using the results of Section 2. Generally, the Higgs contribution is much smaller than that from the $Z$; there is an additional step in the decay chain, so the energies of the neutrinos are smaller. This is especially true if the Higgs is light enough that it decays only to light fermions.

Previously, it was assumed that the gauge bosons were unpolarized, as in Eq. (4.2) [5]. Given our new results, we see that a proper treatment of polarization changes the second moments by no more than 5% for the tensor polarization, and no more than 10% for the longitudinal polarization. In the rest of this Section, we will ignore the small effect of polarization and treat decays of unpolarized gauge bosons only.

The neutrino energy distribution from $W$ or $Z$ bosons at rest is simply a delta function with support at half the gauge-boson mass,

$$\left(\frac{dN}{dE_r}\right)^{\text{rest}}_{WW,ZZ}(E) = \Gamma \delta(E - m_B/2), \tag{4.8}$$

where $m_B$ is the gauge-boson mass and $\Gamma$ is the branching ratio for direct decay into muon neutrinos (this expression should be multiplied by 2 for $Z$ bosons). Applying Eq. (2.1), the moving-frame distribution from unpolarized $W$ or $Z$ bosons moving with energy $E_i$ and velocity $\beta$ is

$$\left(\frac{dN}{dE_\nu}\right)^{\oplus}_{WW,ZZ}(E_i, E_\nu) = \begin{cases} \Gamma(m_B \gamma \beta)^{-1} & \text{for } \frac{E_i}{2}(1-\beta) < E_\nu < \frac{E_i}{2}(1+\beta), \\ 0 & \text{otherwise.} \end{cases} \tag{4.9}$$

Here we have neglected the contribution from other $W$ decay modes. Although the contribution to $dN/dE$ from these other modes may be significant at low energies, these low-energy neutrinos provide little contribution to the upward-muon flux, as discussed above, so the final result obtained using Eq. (4.9) should be accurate to better than 4%.

The energy distribution of neutrinos from gauge-boson decay in the Sun can be obtained by applying Eq. (2.6). The result for the scaled second moment of the neutrino distribution from $W$ decay in the Sun is then [5]

$$\langle Nz^2 \rangle^{\odot}_{WW,i} = \frac{\Gamma_{W \to \mu\nu}}{\beta} \left. \frac{2 + 2E\tau_i(1+\alpha_i) + E^2 \tau_i^2 \alpha_i(1+\alpha_i)}{E_i^3 \tau_i^3 \alpha_i(\alpha_i^2 - 1)(1 + E\tau_i)^{\alpha_i + 1}} \right|_{E = E_i(1-\beta)/2}^{E = E_i(1+\beta)/2}, \tag{4.10}$$

for $W$'s injected into the core of the Sun with energy $E_i$ and velocity $\beta$. Once again, the expression for $ZZ$ pairs is obtained by replacing $\Gamma_{W \to \mu\nu}$ with $2\Gamma_{Z \to \nu_\mu \bar\nu_\mu}$, where the factor of two counts the $Z$ bosons. This does not include the contributions from the decays of the $W$ and $Z$ bosons to heavy fermions, but again, we have already seen that this effect is less than about 4%.

Proper treatment of the effects of vector-boson polarization on the scaled second moment from the Sun leads to a very small correction to Eq. (4.10). In the low-energy limit, the effect of the polarization vanishes, as we have seen. In addition, the effect of the polarization must vanish in the high-energy limit as well. This is because the Lorentz boost factor suppresses the contribution of neutrinos with $\cos\theta \simeq -1$ (antialigned with the gauge-boson direction of motion), but the interactions suppress those with $\cos\theta \simeq 1$ (aligned) because of their higher energy. The net result is that the integrated effect of the polarization is never more than 2%, and is thus negligible for neutrinos from gauge-boson decay in the Sun.

## 5. Neutrinos from Top-Quark Decays

In many cases where the WIMP is heavier than the top quark, WIMPs annihilate predominantly to top quarks, especially if the top is as heavy as recent evidence suggests [22][23]. Therefore, the neutrino spectrum from top-quark decay is needed for a proper calculation of the flux of energetic neutrinos from WIMP annihilation in the Sun or Earth.

In previous work [3][5], the top-quark mass was assumed to be 60 GeV. In this case, calculation of the hadronization and decay channels of the top were quite difficult, and the resulting neutrino spectrum, calculated using Monte Carlo techniques, was accordingly uncertain. In addition, the results were highly dependent on the assumed top-quark mass. Given that the top quark is significantly heavier than the $W$ [22][23], the calculation must be repeated. The top quark is heavy, so it decays almost exclusively to a $W$ boson and a $b$ quark. Hadronization effects are unimportant, and decay is essentially a free-particle weak decay. Therefore, our new results are less subject to theoretical uncertainty.

The neutrino spectrum from top quarks follows from the neutrino spectra of its decay products, the $W$ boson and the $b$ quark, which we have already computed. We will first calculate the neutrino spectra from top-quark decay in the Earth and Sun and assume that the $W$ bosons from top-quark decay are unpolarized. We will then calculate $\langle Nz^2 \rangle$ from top-quark decay in the Earth both with and without polarization and show that polarization of the gauge boson is only a 3% effect. Also, the results of Ref. [13] show that the positive and negative helicity amplitudes for top-quark production are equal. Therefore, the top quarks produced by neutralino annihilation are unpolarized, and we will consider unpolarized top quarks only. We will then present an analytic approximation to $\langle Nz^2 \rangle$ for top-quark decay in the Sun.

First consider the case where the top quark decays at rest. The top decays $t \to Wb$ with a branching ratio close to unity. The spectrum of neutrinos from top-quark decay at rest is given by a sum of the $W$-decay contribution and the $b$-decay contribution:

$$\left(\frac{dN}{dE}\right)^{\text{rest}}_{t\bar{t}}(E) = \left(\frac{dN}{dE}\right)_{WW}(E_W, E) + \left(\frac{dN}{dE}\right)_{b\bar{b}}(E_b, E). \quad (5.1)$$

In the rest frame of the top quark, the $W$ has energy

$$E_W = \frac{m_t^2 + m_W^2}{2m_t}, \quad (5.2)$$

and its velocity is

$$\beta_W = \frac{m_t^2 - m_W^2}{m_t^2 + m_W^2} = \frac{E_b}{E_W}. \quad (5.3)$$

The $W$-boson contribution is given by Eq. (4.9) with injection energy $E_W$. The $b$ quark is first slowed from its injection energy, $E_b = (m_t^2 - m_W^2)/(2m_t)$, by hadronization to an energy $E_d = z_f E_b$, and has a velocity $\beta$ close to unity. The $b$-quark contribution is then given by the $\beta \to 1$ limit of Eq. (3.8).

The boost formula, Eq. (2.1), yields the neutrino spectrum from unpolarized top quarks moving with velocity $\beta_t$ and $\gamma_t = (1 - \beta^2)^{-1/2}$,

$$\left.\frac{dN}{dE}\right|^{\oplus}_{t\bar{t}}(E_\nu) = \frac{\Gamma_{W \to \mu\nu}}{2\gamma_t \beta_t E_W \beta_W} \ln \frac{\min(E_+, \epsilon_+)}{\max(E_-, \epsilon_-)} \Theta\left(\gamma_t(1-\beta_t)\epsilon_- < E_\nu < \gamma_t(1+\beta_t)\epsilon_+\right)$$
$$+ \frac{\Gamma_{b \to \mu\nu X}}{2\gamma_t \beta_t E_d} D_b\left[E_-/E_d, \min(1, E_+/E_d)\right] \Theta\left(E_\nu < \gamma_t(1+\beta_t)E_d\right) \quad (5.4)$$

where $\epsilon_\pm = E_W(1 \pm \beta_W)/2$ with $E_W$ and $\beta_W$ equal to their values in the top-quark rest frame [Eqs. (5.2) and (5.3)], $E_\pm = E_\nu \gamma_t^{-1}(1 \mp \beta_t)^{-1}$, and $\Theta(x) = 1$ if $x$ is true and $\Theta(x) = 0$ otherwise. Also, the $b$-quark decay energy is $E_d = z_f E_b$ where $E_b$ is the energy of the $b$ quark in the top-quark rest frame. The function $D_b$ is given by

$$D_b[x,y] = \frac{1}{3}\left[9(x^2 - y^2) + 4(x^3 - y^3) + 6\ln\frac{y}{x}\right]. \quad (5.5)$$

The kinematics of the decay dictate that the $b$ quarks carry a smaller fraction of the energy than the $W$ bosons, their energy is further degraded by hadronization, and they undergo three-body—rather than two-body—decay. Therefore, even though the $b$-quark contribution to $(dN/dE)^{\text{rest}}_{t\bar{t}}$ may be significant at low energies, it will contribute only about 9% to the final result for the upward-muon flux.

The distribution from the Sun is obtained by applying Eq. (2.6) to Eq. (5.4). It should be noted that application of Eq. (2.6) to Eq. (5.4) does not take into account the stopping of $b$ quarks in the Sun; however, neutrinos from $b$ quarks contribute only 9% of the upward-muon signal, so little accuracy will be lost by neglecting this effect.

We now calculate $\langle Nz^2 \rangle$ for top-quark decay in the Earth, and quantify the effect of gauge-boson polarization on the results for the upward-muon signal. Consider the case where the top quark decays at rest. A fraction [24]

$$f_L = \frac{1}{1 + 2\frac{m_W^2}{m_t^2}}, \quad (5.6)$$

of the $W$ bosons are produced in the the longitudinal-helicity state, and the rest (a fraction $1 - f_L$) are produced in the transverse-helicity states, with equal probability for positive and negative helicity states. The result for $\langle Nz^2 \rangle$ for top quarks at rest ($E_t = m_t$ where $E_t$ is the top-quark energy) then follows from the results of the previous Section. The only subtlety is that the second moment must be scaled by the square of the top-quark mass rather than the square of the $W$-boson energy. The contribution from the semileptonic $b$ decays can also be included. Thus, we have

$$\langle Nz^2\rangle_{t\bar{t}}^{\oplus}(E_t = m_t) = \frac{\Gamma_{W\to\mu\nu}E_W^2}{4m_t^2}\left[f_L\left(1+\frac{\beta_W^2}{5}\right) + (1-f_L)\left(1+\frac{2\beta_W^2}{5}\right)\right]$$
$$+ \Gamma_{b\to\mu\nu X}\frac{2z_f^2 E_b^2}{15m_t^2}$$
$$= \frac{\Gamma_{W\to\mu\nu}E_W^2}{4m_t^2}\left[1+\frac{1}{5}\beta_W^2(2-f_L)\right] + \Gamma_{b\to\mu\nu X}\frac{2z_f^2 E_b^2}{15m_t^2}, \quad (5.7)$$

For $m_t = 174$ GeV [22], $f_L \approx 0.7$ and $\beta_W^2 = 0.4$. If the $W$ bosons were produced unpolarized, then we would have $f_L = 0.33$, so that consideration of the polarization of the $W$ bosons in the decay $W \to \mu\nu$ results in a small change from the unpolarized case, less than about 3%. And as discussed above, the contribution to $\langle Nz^2\rangle$ from $b$ quarks is subdominant (about 9% of the total).

The top quark decays immediately, so the scaled second moment from decay of a top quark in the Earth moving with velocity $\beta_t$ and energy $E_t$ is

$$\langle Nz^2\rangle_{t\bar{t}}^{\oplus}(E_t) = \left(1+\frac{\beta_t^2}{3}\right)\langle Nz^2\rangle_{t\bar{t}}^{\oplus}(E_t = m_t). \quad (5.8)$$

Therefore, the $\langle Nz^2\rangle$ for tops moving with velocity $\beta_t$ is just $(1+\beta_t^2/3)$ times that for top quarks at rest, as given in Eq. (5.7).

Eqs. (5.7) and (5.8) provide expressions for the scaled second moment of the neutrino distribution from top quarks from WIMP annihilation in the Earth, accurate to at least 3%.

In order to obtain $\langle Nz^2\rangle$ from top-quark decay in the Sun, the stopping and absorption of neutrinos, as given in Eqs. (2.4) and (2.5), and stopping of the $b$-quark hadrons must be included.

Neutrinos come predominantly from the $W$ boson produced by the top-quark decay, and to a lesser extent from the $b$ quarks. We already know the $\langle Nz^2\rangle$ from injection of these particles in the center of the Sun. The only catch is that, to use Eq. (4.10), we must ignore polarization of the $W$; the results of the previous subsection show that $W$ polarization in top decay is less than a 3% effect. Therefore, the easiest way to obtain the result for $\langle Nz^2\rangle$ from top quarks from the Sun is to integrate the $\langle Nz^2\rangle$ for $W$ bosons and $b$ quarks over the appropriate injection energies.

The expression for $\langle Nz^2\rangle$ as a function of the top-quark energy, $E_t$, for top quarks injected into the Sun follows from Eq. (2.12) and is

$$\langle Nz^2\rangle_{t\bar{t}}^{\odot}(E_t) = \frac{1}{E_t^2}\sum_{f=b\bar{b},WW}\frac{1}{2\gamma_t E_f \beta_f \beta_t}\int_{\gamma_t E_f(1-\beta_t\beta_f)}^{\gamma_t E_f(1+\beta_t\beta_f)} E^2\langle Nz^2\rangle_f^{\odot}(E)\,dE, \quad (5.9)$$

where we have included the sum over both the $W$ bosons and $b$ quarks from top-quark decay. The integral is over the injection energy of the decay particles, and the moments $\langle Nz^2\rangle_f^{\odot}$ include the effects of interactions with the solar medium. For $W$ bosons, $\langle Nz^2\rangle^{\odot}$ is given by Eq. (4.10), and for $b$ quarks by Eq. (3.15). The quantities $E_t$, $\beta_t$, and $\gamma_t$ are the energy, velocity, and $\gamma$ factors of the top quark, and $E_W$ and $\beta_W$ are the energy and velocity of the $W$ boson in the rest frame of the top, as given in Eqs. (5.2) and (5.3). The quantities $E_b = (m_t^2 - m_W^2)/(2m_t)$ and $\beta_b \simeq 1$ are the energy and velocity of the $b$ quark in the rest frame of the top. Also, recall that $\langle Nz^2\rangle$ for neutrinos from the Sun are different than those for antineutrinos.

The results for $\langle Nz^2\rangle$ for neutrinos and antineutrinos from injection of top quarks into the core of the Sun and Earth are shown in Fig. 1. There are three bands of curves: the top is the result for $\langle Nz^2\rangle$ from the Earth (and is the same for both neutrinos and antineutrinos). The center band is the result for antineutrinos from top quarks injected into the core of the Sun, and the lower band is that for neutrinos from top quarks injected into the core of the Sun. The curves for antineutrinos from the Sun are higher than those for neutrinos since neutrinos interact more strongly with the solar medium than antineutrinos and are therefore more strongly attenuated as they pass through the Sun. Comparison of the results for $\langle Nz^2\rangle$ from the Sun and Earth illustrates that absorption and stopping of neutrinos in the Sun are very important effects.

In each band, the solid curve is the total result for $\langle Nz^2\rangle$ which accounts for neutrinos from both the $W$ boson and $b$ quark. In each case we have taken $m_t = 174\,\mathrm{GeV}$ [22]. The dashed curve is the result obtained ignoring the contribution from $b$ quarks. For neutrinos from the Sun, the $b$-quark contribution is about 10% at low energies, and increases to roughly 25% at energies around 1000 GeV. The increased importance of the $b$-quark neutrinos from the Sun at higher energies is due to the fact that the neutrinos from $W$ bosons are more energetic and therefore more strongly absorbed as they travel through the Sun.

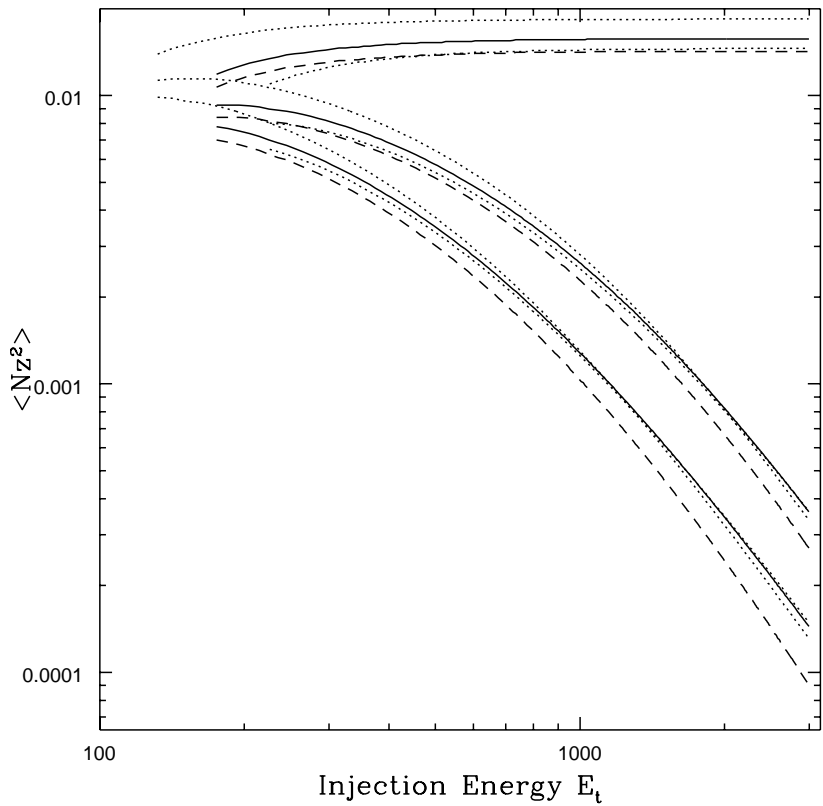

Fig. 1. Plot of $\langle Nz^2 \rangle$ for top quarks injected into the core of the Sun and Earth as a function of the injection energy $E_t$. See text for a description of the curves.

In Fig. 1, we also illustrate the effect of varying the top-quark mass. The dotted curves above and below the solid curves in each band are the total $\langle Nz^2 \rangle$ obtained using $m_t = 131$ GeV (a conservative lower limit [23]) and $m_t = 225$ GeV (the $3\sigma$ upper limit suggested by the CDF data [22]), respectively.

We have crafted the following analytic fit which provides a good approximation (better than 10%) to the $\langle Nz^2 \rangle$ over the range $m_t \leq E_t \leq 3000$ GeV for neutrinos from the Sun (the two lower solid curves curves in Fig. 1):

$$\log_{10}\left[\langle Nz^2\rangle^\odot_{t\bar{t},i}\right] = A_i(\log_{10} E_t)^2 - B_i(\log_{10} E_t) + C_i. \qquad (5.10)$$

For neutrinos, $A_\nu = -0.825$, $B_\nu = -3.31$, and $C_\nu = -5.39$. For antineutrinos, $A_{\bar\nu} = -0.889$, $B_{\bar\nu} = -3.94$, and $C_{\bar\nu} = -6.40$. Of course, if the top-quark mass differs from 174 GeV, these coefficients will change slightly.

## 6. Summary

In this paper we have calculated the differential energy spectra of muon neutrinos and antineutrinos from particle decay in the Sun and Earth. These results will be useful for computing the rates for upward-muon events in neutrino telescopes from annihilation of WIMPs that have accumulated in the Sun and/or Earth. The analytic expressions we have derived for neutrino spectra from decay of particles injected in the Sun result from the two- or three-body decay kinematics and include the effects of hadronization. The results for particle decay in the Sun take into account the additional effects of stopping of heavy hadrons in the Sun and stopping and absorption of neutrinos as they pass through the Sun. Neutrino spectra are given for all the WIMP-annihilation channels that will produce energetic neutrinos. These expressions will give results for the flux of neutrino-induced upward muons accurate to better than $\mathcal{O}(10\%)$ and can be used in place of Monte Carlo calculations.

In particular, the energy spectra from injection of $\tau\bar\tau$, $c\bar c$, and $b\bar b$, pairs in the Earth are given by Eqs. (3.2), (3.8), and (3.11), respectively. The neutrino spectra from injection of $W^+W^-$ and $ZZ$ pairs in the Earth are given by Eq. (4.9), and the spectrum from injection of top quarks in the Earth is given by Eq. (5.4). The antineutrino spectra from particle decays in the Earth are the same as the neutrino spectra. The neutrino spectra from injection of $\tau\bar\tau$, $W^+W^-$, and $t\bar{t}$ pairs in the Sun are obtained by applying Eq. (2.6) to Eqs. (3.2), (4.9), and (5.4), respectively, and the neutrino spectra from injection of $b\bar{b}$ and $c\bar{c}$ pairs in the Sun are given by Eqs. (3.9) and (3.12), respectively. Recall that interactions of antineutrinos with the solar medium differ than those of neutrinos, so the antineutrino and neutrino spectra from particle decay in the Sun differ. The neutrino spectra from decays of Higgs bosons in the Sun and Earth can be obtained from the spectra of the Higgs decays products by applying Eq. (2.11).

The probability of detecting a neutrino-induced upward muon is proportional to the square of the neutrino energy; one power comes from the charged-current cross section for muon production, and the other comes from the muon range. We argued that in most cases, the detector thresholds are small enough compared to the WIMP mass that the thresholds can be ignored. If so, then the upward-muon flux from WIMP annihilation is proportional to the second moment of the neutrino energy distribution, Eq. (1.2) will provide an accurate estimate of the muon flux from a given WIMP candidate, and detailed information about the shape of the neutrino energy distribution is not required. For this reason, we have provided analytic expressions for the scaled second moments of the neutrino distribution, $\langle Nz^2 \rangle$, for each WIMP-annihilation channel. The scaled second moments from injection of $\tau\bar{\tau}$, $b\bar{b}$, $c\bar{c}$, $W^+W^-$, $ZZ$, and $t\bar{t}$ pairs in the Earth are given by Eqs. (3.3), (3.13), (3.14), (4.5), (4.6), and (5.8), respectively. The scaled second moments from injection of $\tau\bar{\tau}$ and $t\bar{t}$ pairs in the Sun are given by Eqs. (3.4) and (5.10), respectively. Those from injection of $b\bar{b}$ and $c\bar{c}$ pairs in the Sun are given by Eq. (3.15), and those from injection of $W^+W^-$ and $ZZ$ pairs are given by (4.10). The scaled second moments from Higgs-boson decay in the Sun and Earth can be obtained from those of the Higgs decay products using Eq. (2.12).

The results for the second moments, $m_{\tilde{\chi}}^2 \langle Nz^2 \rangle$, for neutrinos and antineutrinos from injection of particle-antiparticle pairs in the Sun with energy equal to the WIMP mass, $m_{\tilde{\chi}}$, are plotted in Fig. 2 and Fig. 3, respectively. The strongest signals are from those particles that decay immediately and directly to neutrinos, the $\tau\bar{\tau}$ and gauge-boson final states, and the signal from top quarks is only slightly weaker. The turnover in the gauge-boson signal strength is due to absorption of high-energy neutrinos in the Sun; the signals from other particles will turn over similarly at larger energies. The Figures illustrates the importance of hadronization and stopping of heavy hadrons: the signals from $b$ and $c$ quarks are significantly smaller than those from particles that do not from hadrons.

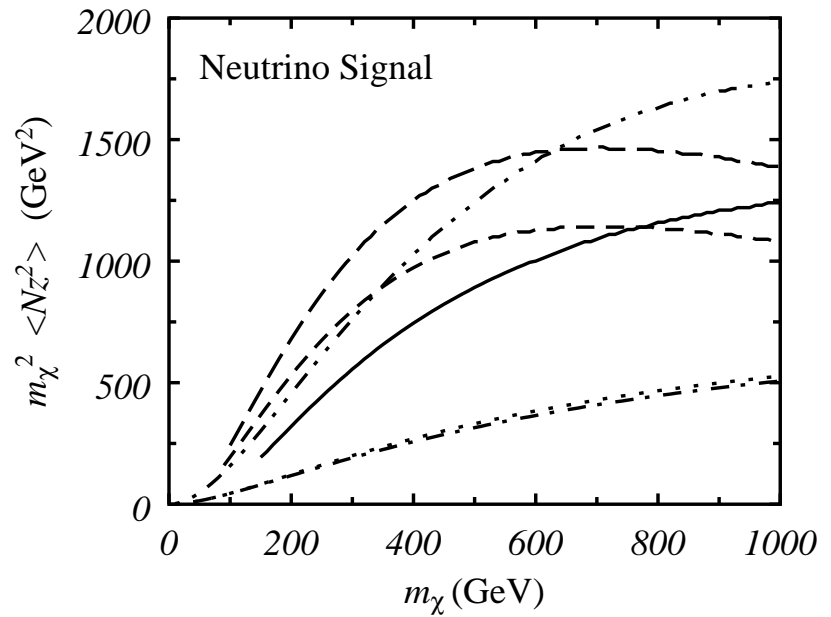

Fig. 2. The second moment, $m_{\tilde{\chi}}^2 \langle Nz^2 \rangle$, of the neutrino energy distribution from injection of particles with energy equal to the WIMP mass $m_{\tilde{\chi}}$ in the Sun. The solid curve is for $t\bar{t}$ pairs, the upper (lower) dashed curve is for $W^+W^-$ ($ZZ$) pairs, and the upper dot-dash curve is for $\tau\bar{\tau}$ pairs. At the bottom are the $b\bar{b}$ and $c\bar{c}$ curves, and the $b\bar{b}$ curve is slightly higher than the $c\bar{c}$ curve.

The results listed here contain the effects of hadronization of heavy quarks and interactions of heavy hadrons and neutrinos with the solar medium. We have included only the neutrinos from prompt particle decay. In some cases, there will be additional neutrinos from secondary decays, but their contribution to the upward-muon flux will be negligible. For example, there will be neutrinos produced by gauge-boson decay to $\tau$ leptons and $b$ and $c$ quarks in addition to the prompt neutrinos from gauge-boson decay, but we estimate that their contribution to the upward-muon flux will be less than 4% of the total. Also, there will be additional neutrinos from the $c$ quarks produced by $b$ decay, but we estimate that these will contribute roughly 10% of the total upward-muon flux. The additional contribution to the upward-muon flux from decays of $\tau$ leptons produced by charged-current interaction of $\tau$ neutrinos as they pass through the Sun are negligible. If our neutrino spectra are used to compute event rates for detection techniques which are proportional to the first moment of the neutrino energy distribution (such as searches for contained events), then the contribution from the neutrinos from secondary decays which we neglect may be significant and the event rates will be underestimated.

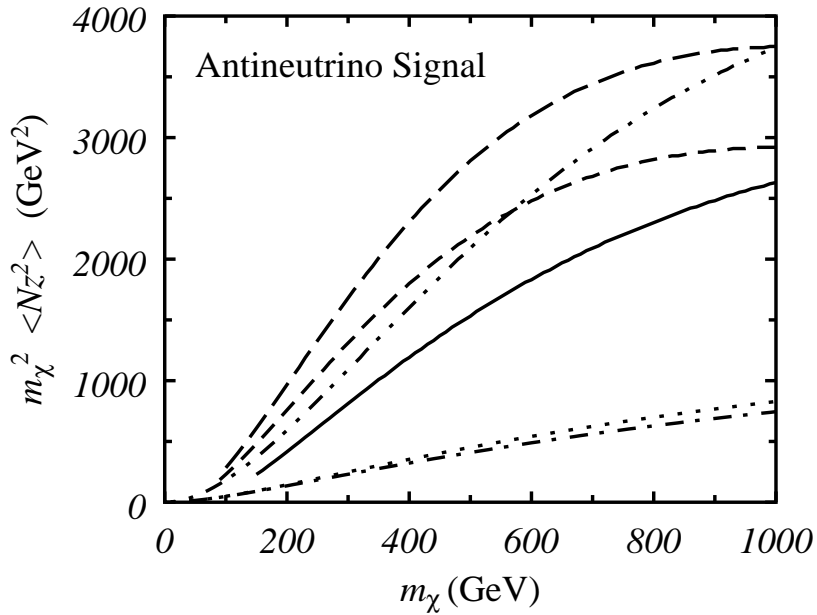

Fig. 3. Same as Fig. 2, but for antineutrinos.

The neutrino energy spectrum from decay of a given particle depends on the energy of the decaying particle, and it may depend on the polarization state as well. We have evaluated the effect of polarization on the upward-muon flux; we found that it is never more than a 10% effect and can typically be accounted for easily

The largest theoretical uncertainty involves modeling the effect of stopping of heavy hadrons in the Sun. This can be traced to the uncertainty in the values adopted for the stopping coefficient, $E_c$ [c.f. Eq. (2.7)]. However, neutralino annihilation into $b$ and $c$ quarks is always accompanied by annihilation into $\tau$ leptons (with a branching ratio enhanced by radiative corrections [25]). If heavy enough, WIMP annihilation is predominantly to $t\bar{t}$ pairs and possibly gauge-boson final states, so the uncertainty in the total upward-muon flux due to uncertainty in the stopping effect is diluted. For all annihilation channels, there is also the uncertainty in the absorption and stopping length for neutrinos and antineutrinos in the Sun, but this is small at energies near a TeV, and is virtually negligible at lower energies. These uncertainties do not effect the neutrino spectra from the Earth. Note that these theoretical uncertainties must also enter Monte Carlo calculations and are not introduced by our analytic treatment.

## 7. Acknowledgments

M.K. was supported by the U.S. Department of Energy under contract DE-FG02-90ER40542, and G.J. was supported by DE-FG02-85ER40231 at Syracuse University.